\documentclass[a4paper,11pt]{article}
\usepackage{jheppub} % for details on the use of the package, please see the JINST-author-manual

\preprint{MS-TP-24-36}

\title{\boldmath Hadronic decay of vector charmonium from the lattice}

\author[a]{Beno{\^i}t Blossier\orcidlink{0000-0001-5620-6006},}
\author[b]{Jochen Heitger\orcidlink{0000-0001-5181-9145},}
\author[a,b]{Jan Neuendorf\orcidlink{0000-0001-6177-7014},}
\author[a,c]{and Teseo San José\orcidlink{0000-0001-5402-2633}}

\affiliation[a]{
	Laboratoire de Physique des 2 Infinis Irène Joliot-Curie,
	CNRS/IN2P3,\\
	Université Paris-Saclay,
	91405 Orsay Cedex,
	France
}
\affiliation[b]{
	Institut f{\"u}r Theoretische Physik,
	Universit{\"a}t M{\"u}nster,\\
	Wilhelm-Klemm-Str. 9,
	48149 M{\"u}nster,
	Germany
}
\affiliation[c]{
    Higgs Centre for Theoretical Physics, School of Physics and Astronomy,\\
    The University of Edinburgh, Edinburgh EH9 3FD, United Kingdom
}

\emailAdd{benoit.blossier@ijclab.in2p3.fr}
\emailAdd{heitger@uni-muenster.de}
\emailAdd{j\_neue03@uni-muenster.de}
\emailAdd{san-jose-perez@ijclab.in2p3.fr}

\abstract{Estimating decay parameters in lattice simulations is a computationally demanding problem, requiring several volumes and momenta.
We explore an alternative approach, where the transition amplitude can be extracted from the spectral decomposition of particular ratios built from correlation functions.
This so-called ratio method has the advantage of not needing various irreducible representations or volumes, and it allows us to predict the decay width $\Gamma$ and the energy shift $\epsilon$ of the spectrum directly.
In this work, we apply this method to study the hadronic decay $\HepProcess{\Pgya\to\APD\PD}$ on two \acs{cls} $N_\text{f}=2$ ensembles. This approach requires close to on-shell kinematics to work, and we employ \aclp{tbc} to precisely tune the on-shell point.
Although our study is yet to approach the continuum limit, we find a value of $\Gamma$ fully compatible to the physical result, and $\epsilon$ informs us by how much our spectrum would shift in a fully dynamical simulation.
Besides lattice calculations, many analytical tools have been proposed to understand decay processes. A relatively simple, early example is the ${}^3P_0$ quark model, which provides a physical insight of the decay process.}

\begin{document}
\maketitle
\flushbottom

\section{Introduction} \label{sec:introduction}

% What are we doing?

In this paper, we use lattice simulations to study the strong decay $\HepProcess{\Pgya\to\APD\PD}$, its decay width, and the energy shift between the initial and final states. This is a transition between the second excited state of vector charmonium to two pseudo-scalar $\PD$ mesons in a $p$-wave configuration. It is the main decay channel of the $\Pgya$ resonance, accounting for almost its entire branching ratio \cite{ParticleDataGroup:2022pth}, and it takes place close to threshold. Given the small phase space allowed and the large meson masses, the final $\APD\PD$ state is non-relativistic.

% Why are we doing it?

In general, the mass spectrum of hadrons and their decays through strong interactions are fundamental subjects of experiment. Comparing them to their theoretical prediction helps us understand the basic properties of hadrons, and given their non-perturbative nature, \ac{lqcd} is in a privileged position to compute these quantities from first principles.
There are a number of technical difficulties one must overcome to obtain lattice predictions, however: There are no scattering states in the finite volume of an Euclidean simulation; the momenta on the lattice are quantized, such that the decay channel may not even be possible on-shell, depending on the value of the lattice spacing and the quark masses; and the large computational cost makes giving predictions in the continuum challenging.

% Why are we doing it? What is the novelty?

This is exploratory work, where we study alternatives to treat these issues. As we explain in the following in more detail, we study the decay employing ratios of correlation functions, and we predict the energy shift of $\Pgya$ and $\APD\PD$ considering a two-level system for different values of the meson momentum. The latter is fixed using \acp{ptbc}, such that we can tune the on-shell point precisely, or explore the off-shell behavior of the matrix element at will. In addition, we attempt to minimize the computational costs of the more standard methods by extracting the hadronic transition amplitude directly from ratios of three-point and two-point functions. This avoids the need to use several lattice volumes to predict the scattering properties. Finally, we examine the ${}^3P_0$ quark model \cite{MICU1969521,Carlitz:1970xb}, an analytic tool developed in the late 60's and early 70's, and study how well it can describe the data of a modern lattice calculation.

% What alternatives there are?

The main lattice method to study the decay processes was initially proposed in \cite{Luscher:1985dn,Luscher:1986pf,Luscher:1991cf,Luscher:1990ux}. It considers that, as the simulation volume is increased, the two-body interaction in the final state will approach the non-interacting limit. Then, it provides a relation between the energy spectrum in finite volume and the scattering phase shift in infinite volume, which holds as long as the interactions occur below the inelastic threshold. This method has been used in \cite{Lang:2015sba,Piemonte:2019cbi} to compute the same quantities that we study in this paper.

However, other methods exist which try to avoid the large computational costs associated with the analysis of several volumes. Introduced in \cite{Pennanen:2000yk} by Michael and Pennanen, the ratio method used here relies on the narrow width approximation to extract the mixing between the initial and final states from ratios of correlation functions.
Their original motivation was to study the string breaking in heavy mesons, where an initial state of static quarks $Q\bar{Q}$ decays to a pair of static-light mesons. Particularly interesting for them were decays close to threshold, as string breaking becomes especially relevant.
Their formalism has been applied to studying the mixing between scalar glueballs and flavour-singlet mesons \cite{McNeile:2000xx}, the decay of $J^{PC}=1^{-+}$ spin-exotic hybrids \cite{McNeile:2002az}, $\HepProcess{\Prho\to\Ppi\Ppi}$ \cite{McNeile:2002fh}, as well as $\HepProcess{\PB(0^+)\to\PB\pi}$, $\HepProcess{\PBs(0^+)\to\PB\PK}$ \cite{McNeile:2004rf}, \HepProcess{B_0^*\to\PB\Ppi} and \HepProcess{B_1^*\to\PB_0^*\Ppi} \cite{Blossier:2014vea}.
More recently, it has been extended to baryonic strong decays \cite{Alexandrou:2013ata,Alexandrou:2015hxa}, with special success for $\HepProcess{\PDelta\to\Ppi\PN}$. While these works study the transition between ground states with different quantum numbers, our work generalizes the expressions of \cite{McNeile:2002fh} for a transition from an excited state.

In addition to the lattice approach, there are also phenomenological methods to predict the decay width. The K-matrix framework \cite{https://doi.org/10.1002/andp.19955070504} models the amplitudes of $2 \to \text{hadron} \to 2$ scatterings and $\text{hadron} \to 2$ decays. It preserves the unitarity and analyticity of the underlying scattering amplitudes and can be formulated in a Lorentz invariant fashion \cite{ParticleDataGroup:2022pth}. This method has been used in \cite{Shamov:2016mxe,Uglov:2016orr,Coito:2017ppc,Hanhart:2023fud} to predict the mass and decay width of $\Pgya$. Following the discussion in \cite{Hanhart:2023fud}, the free parameters of the K-matrix, which account for multiple decay channels and effects, must be fitted using the experimental cross section of, in this case, $\HepProcess{\Ppositron\Pelectron\to\text{hadrons}}$ in the range of energies around the open-charm decay channel.
In particular, we compare our results with another analytic approach, the ${}^3P_0$ quark model \cite{MICU1969521,Carlitz:1970xb}. It considers the two initial charm quarks as spectators of the decay, such that their quantum numbers and momenta are conserved. To form the final state, a quark-antiquark pair is created out of the vacuum, and the quarks and their quantum numbers are then rearranged. The model in its simplest incarnation has only one free parameter, the coupling $\beta$ associated to the creation of the quark-antiquark pair. Beyond the general concept, there are several implementations, depending on how the meson wave functions are described, or if a relativistic treatment is used or not. Given the characteristics of our problem, we follow the non-relativistic work in \cite{LeYaouanc:1972vsx}, where the harmonic oscillator in spherical coordinates is used to describe the meson wave functions.

This manuscript is organized as follows: \Cref{sec:methodology} explains how we extract the transition matrix element $\braket*{\APD\PD}{\Pgya}$. \Cref{sec:lattice-setup} details the lattice simulations, including our use of \ac{ptbc} and the interaction between them and the various correlators. \Cref{sec:analysis} presents our data analysis and gathers our results for the decay width and energy shifts. \Cref{sec:conclusions} gathers our conclusions and outlook for the future, and \cref{sec:quark-model} explains the ${}^3P_0$ quark model.
\section{Methodology} \label{sec:methodology}

\begin{figure}
    \centering
    \begin{subfigure}[t]{0.49\textwidth}
        \centering
        \includegraphics[scale=1]{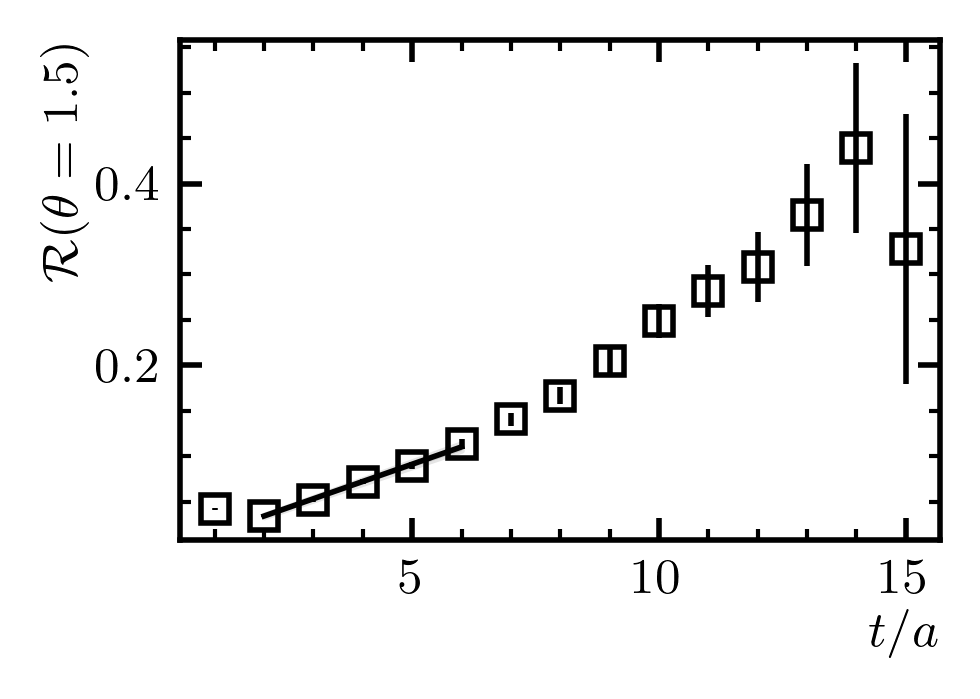}
    \end{subfigure}
    \begin{subfigure}[t]{0.49\textwidth}
        \centering
        \includegraphics[scale=1]{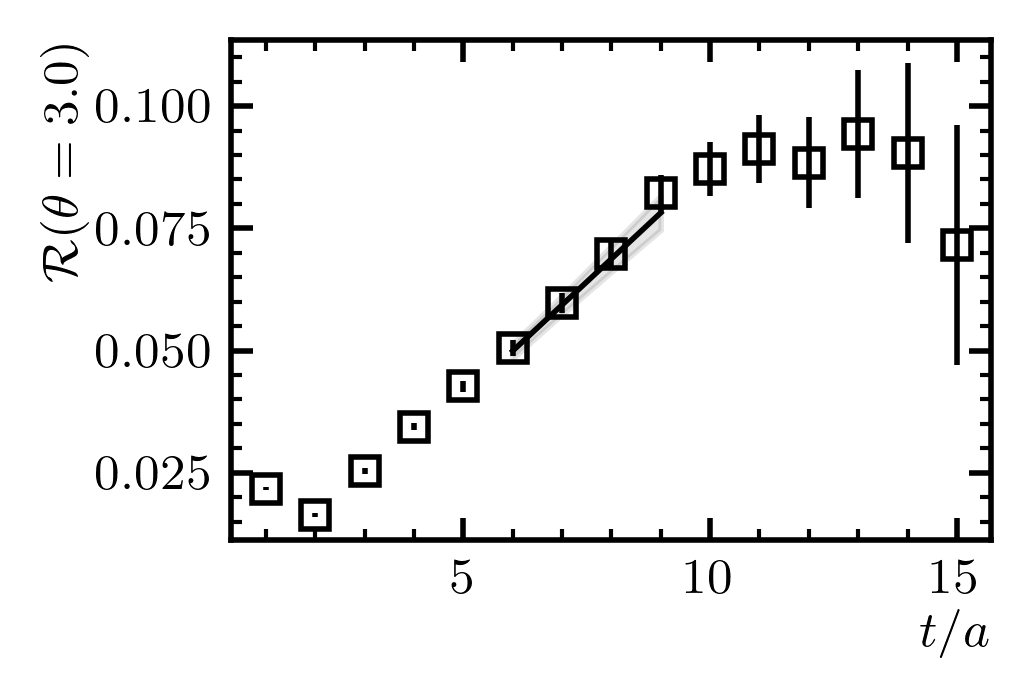}
    \end{subfigure}
    \caption{Fit of the ratio \cref{eq:off-shell-ratio} on ensemble D5 at twist angle $\theta=1.5$ (LEFT), and ensemble E5 at twist angle $\theta=3$ (RIGHT).}
    \label{fig:fit-slope}
\end{figure}

The basis of the method presented in \cite{Pennanen:2000yk,McNeile:2000xx} is a careful analysis of the spectral decomposition for the relevant matrix elements. In a certain regime, it is possible to isolate the hadronic matrix element describing the mixing between $\Pgya$ and $\APD\PD$ thanks to a time enhancement. In this work, we need to consider a mixing between the second excited state of vector charmonium, and the ground state of a $\APD\PD$ system in a $p$-wave configuration (see the relevant diagrams in \cref{sec:lattice-setup}). In this situation, one finds that some of the original expressions to extract the hadronic mixing \cite{McNeile:2002fh} are modified. The relations in this section are derived within the following setup: The state $\Pgya$ is isolated with a \ac{gevp}, but not the $\APD\PD$ pair; and the lattice spacing is sufficiently fine to replace sums by integrals over time. We consider operators $O^{\psi}_i$ to interpolate the charmonium states, and one operator $O^{\APD\PD}$ to reproduce the ground state of two $\PD$-mesons with back-to-back momentum. The basic objects are the propagators of these states,
\begin{equation}
    P^{\psi}_{ji}(t) \equiv \expval{O^{\psi}_j(t) O^{\psi\dagger}_i},
    \quad
    P^{\PD\PD}(t) \equiv \expval{O^{\APD\PD}(t) O^{\APD\PD\dagger}}.
\end{equation}
Organizing the charmonium correlators in a matrix, we can solve the \ac{gevp}
\begin{equation}
    P^{\psi}(t) \bar{v}_{\alpha} = \lambda_{\alpha}(t,t_0) P^{\psi}(t_0) \bar{v}_{\alpha},
    \qquad\qquad
    \lambda_{\alpha}(t,t_0) = e^{-E_{\alpha}(t-t_0)}.
\end{equation}
Knowing the eigenvalues $\lambda_3(t,t_0)$ and eigenvectors $\bar{v}_3$, we can build the matrix element
\begin{equation}
    \bar{T}_3(t) \equiv \bar{v}_{3i}^\dagger T_i(t)
    \quad\text{with}\quad
    T_i(t) = \expval{O^\psi_i(t) O^{\APD\PD}}.    
\end{equation}
To extract the hadronic mixing, one now expands the correlator $\bar{T}_3(t)$ in the Schr\"odinger picture and introduces two complete sets of states at a time $t_i$ when the transition $\HepProcess{\Pgya\to\APD\PD}$ takes place: one set of charmonium states, another of $\APD\PD$ mesons. Since this may occur at any time slice, the correlator is the sum over all possible values for $t_i$. The expression can be simplified using the fact that only $\Pgya$ has an overlap with our charmonium operator, and replacing the summation over time by an integral. This leads to the expression
\begin{equation}
    \bar{T}_3(t) = e^{E_3t_0/2} \sum_{\beta=1}^M x_{3\beta} \mel*{0}{O^{\APD\PD}}{\beta}
    \begin{cases}
        \dfrac{e^{-tE_3}-e^{-tR_\beta}}{R_\beta-E_3}, & E_3 \neq R_\beta,
        \\
        te^{-tE_3}, & E_3 = R_\beta.
    \end{cases}
\end{equation}
We sum over the tower of $\APD\PD$ states, but one can remove this summation by performing a \ac{gevp} also for the final states. We denote the charmonium energies by $E_\alpha$ and the $\APD\PD$ levels by $R_\beta$. We observe that there are two possibilities, depending on whether or not the transition is on-shell. If $E_3 = R_1$,
\begin{equation}
    \label{eq:on-shell-ratio}
    \mathcal{R}(t) \equiv \frac{\abs{\bar{T}_3(t,t_0)}}{\sqrt{P^{\PD\PD}(t)\lambda_3(t,t_0)}} = \abs{x_{31}} t + A,
\end{equation}
where $x_{31} = \braket{\APD\PD}{\Pgya}$ and $A$ is a constant that accounts for other transitions. The denominator cancels a global exponential dependence and normalizes the matrix element. Then, we conclude that the hadronic transition has an enhancement proportional to $t$ with respect to other states, and forming this ratio allows one to extract $x_{31}$ at sufficiently long times. We are interested not only in the on-shell condition, but we want to study the dependence of the matrix element on the $\PD$-meson momentum, venturing into off-shell kinematics. Choosing the non-degenerate scenario, where there is no energy conservation in the process, and denoting the energy difference by $\Delta \equiv (E_3-R_1) / 2$, the ratio is modified to
\begin{equation}
    \label{eq:off-shell-ratio}
    \mathcal{R}(t) = \frac{ \abs{\bar{T}_3(t,t_0)} }{ \sqrt{P^{\PD\PD}(t) \lambda_3(t,t_0)} }
    = \frac{ \abs{x_{31}} }{ \Delta } \sinh(t\Delta) + Ae^{-t\Delta}.
\end{equation}
We observe that for $\Delta=0$ we recover the degenerate scenario. As pointed out in \cite{McNeile:2000xx}, it should also be possible to extract the same hadronic matrix element from the box diagram of $\HepProcess{\APD\PD\to\APD\PD}$ (see \cref{fig:propagator-dd-connected-u-annihilation}). Unfortunately, our situation is more unfavourable. From an intuitive point of view, the $\APD\PD$-system propagates until a time $t_1$, where it annihilates and forms charmonium.
Further, this intermediate, off-shell charmonium propagates until a time $t_2$, when it reconstitutes the $\APD\PD$ pair. In this situation, we cannot prevent the intermediate state to have a coupling to $\PJgyi$, $\Pgyii$, or any other charmonium state for that matter. The problem cannot be avoided, as the signal of the first two charmonia levels is exponentially enhanced with respect to $\Pgya$. Focusing on the situation when the initial and final states are the same, and $R_1 = E_3$, we can follow similar steps as previously done for the triangle diagram. We find
\begin{equation}
  P^{\PD\PD,\text{b}}
  =
  P^{\PD\PD}
  \Bigg(
    \frac{\abs{x_{31}}^2}{2}t^2
    + \frac{\abs{x_{11}}^2}{\Delta_1^2} \big( -1-\Delta_1 t+e^{t\Delta_1} \big)
    + \frac{\abs{x_{21}}^2}{\Delta_2^2} \big( -1-\Delta_2 t+e^{t\Delta_2} \big)
  \Bigg),
\end{equation}
where $\Delta_1 \equiv R_1-E_1 > 0$, $\Delta_2 \equiv R_1-E_2 > 0$, $E_1$ is the energy of $\PJgyi$, and $E_2$ is the energy of $\Pgyii$. In practice, the extra exponential terms, not present in a ground-to-ground transition, prevent us from extracting the mixing $x_{31}$ with the last relation, as the term is no longer enhanced in time.
Besides these expressions, other ratios are given in \cite{McNeile:2000xx}, which assume that both correlators perfectly isolate the relevant state but include the effects from the quark source,
\begin{equation}
    P^{\psi}(t) = v_3 e^{-tE_3}v_3^{\dagger} + v_s \delta_{t0} v_s^{\dagger},
    \quad
    P^{\PD\PD}(t) = w_1 e^{-tR_1}w_1^{\dagger} + w_s \delta_{t0} w_s^{\dagger},
\end{equation}
with $v$ and $w$ two-point function amplitudes. This is not exactly our case, since we do not solve a \ac{gevp} for the $\APD\PD$-system. Yet, it is interesting to test the practical applicability of the expressions. Redoing the same steps as earlier, but using the simplified spectral decomposition given above, we find the same result as in \cite{McNeile:2000xx},
\begin{equation}
    \label{eq:xt}
    x_{\text{T}}(t) \equiv
    \dfrac{\bar{T}_3}{\sqrt{\lambda_3(t,t_0) P^{\PD\PD}}}
    \dfrac{\lambda^{t/2}}{1+\lambda+\dots+\lambda^t} \underset{t \gg 1}{\to} x_{31},
\end{equation}
with $\lambda\equiv \exp(E_3-R_1)$. Finally, a similar expression to $x_{\text{T}}$ can be found using the box diagram \cite{McNeile:2000xx}. However, the necessary assumptions are not fulfilled in our situation, for the same reasons that the box diagram alone does not work. \Cref{fig:fit-slope} shows an example of how \cref{eq:on-shell-ratio,eq:off-shell-ratio} look like in practice on our two lattice ensembles close to on-shell kinematics. Both cases have been fitted using \cref{eq:off-shell-ratio}, but it is clear that $\Delta \to 0$. The slope of the straight line corresponds to the matrix element $\abs{x_{31}}$. Using the ratio $x_{\text{T}}(t)$, we tend to observe short plateaus, and we found that this formula is not as reliable in our situation as \cref{eq:on-shell-ratio,eq:off-shell-ratio}. That is why we will not use it to extract the decay width, although we show some intermediate steps in \cref{fig:mixing-vs-twist}.

% Summarize the physical idea to relate the experiment and the lattice simulation.
To understand how to connect our lattice calculation to the physical decay, we employ non-relativistic quantum mechanics.
% Hamiltonian
We divide the \acs{qcd} Lagrangian in a main diagonal piece $H_0 = \ket{\APD\PD} \bra{\APD\PD} + \ket{\Pgya} \bra{\Pgya}$, and a perturbation that allows the transition to occur, $H_1 = \ket{\APD\PD} \bra{\Pgya} + \ket{\Pgya} \bra{\APD\PD}$. For this assumption to hold, the off-diagonal matrix elements need to be small compared to the unperturbed Hamiltonian.
% Solution to the Schr\"odinger equation.
In the experiment, we need to consider that the momentum of the final state forms a continuum, and we can describe the situation using the saddle-point approximation \cite{Weisskopf1930,Cohen-Tannoudji:101367,Sakurai:2011zz} to solve the Schr\"odinger equation. This method yields the fundamental properties of a decay: the initial state disappears exponentially fast in time as a function of the decay width $\Gamma$; the momenta of the $\APD\PD$ final state follow a Breit-Wigner distribution at sufficiently long times; and the existence of the decay channel modifies the energy of both the initial state and the center of the final distribution. The decay width of the initial state is given by Fermi's golden rule,
\begin{equation}
    \Gamma = 2\pi \int \dd{\Omega} \rho(E=E_i) \abs{x_{31}(E=E_i)}^2,
\end{equation}
where $\Omega$ is the solid angle, $\rho$ is the density of states, and $x_{31} = \braket*{\APD\PD(E,\bar{p})}{\psi(3770)}$ gives the transition amplitude. When the final state is non-interacting, such that each $\PD$-meson can be approximated by a plane wave, and the $\PD$-mesons cannot be distinguished, one can extract a useful relation between the lattice calculation and the decay width \cite{McNeile:2002az,McNeile:2002fh},
\begin{equation}
    \label{eq:decay-width-lattice}
	\Gamma = \frac{L^3}{24\pi} p_i E_i \abs{x_{31}(E=E_i)}^2
	\quad \text{with} \quad p_i^2 = \frac{E_i^2}{4} - m_{\PD}^2.
\end{equation}
The apparent volume dependence of $\Gamma$ has to cancel with the behavior of the matrix element. We expect this cancelation to be exact for sufficiently large volumes. The expression can also be generalized to a virtual transition to ease the comparison with other theoretical calculations, although only an on-shell decay width has a physical meaning. Regarding the energy shift, it is much easier to extract it looking at the lattice simulation, where we have a two-level system with discrete momentum and (possibly) degenerate states. According to quantum mechanics, this system yields Rabi's formula and an oscillation between the two levels, $\Pgya$ and $\APD\PD$, rather than a decay. The basic object in this situation is a two-level transfer matrix, whose eigenvalues predict the energy shift of the original states $\Pgya$ and $\APD\PD$ due to their mixing,
\begin{equation}
    \label{eq:transfer-matrix}
	T =
	\begin{pmatrix}
		-\delta/2 & x_{31} \\
		x^*_{31}  & \delta/2
	\end{pmatrix},
\end{equation}
where $\delta$ is the energy difference between $\Pgya$ and $\APD\PD$. In solving the eigenvalue problem, the hadronic mixing separates the initial spectrum by an amount $\epsilon^2 = \abs{x_{31}}^2+\delta^2/4$. When the transition is on-shell, $\delta=0$, and the matrix element makes the states non-degenerate. An interesting note is that the energy shift appears as a first-order perturbation on the two-level setup, while it is second-order in the saddle-point approach.

% Range of applicability.

Despite its advantages, this approach makes a series of assumptions that are not true in general, and one should be careful to apply it under the correct circumstances. First, the method does not attempt to connect the finite-volume matrix elements to an infinite-volume process. This is important because no scattering may occur on the lattice. A resonance appears as a pole in the scattering amplitude, with the decay width indicating the distance from the real axis. Therefore, one expects the systematic error of this method to be reduced in the narrow-width approximation with a physical tuning of masses and momenta. Second, the method assumes that the resonance is an asymptotic state, but different techniques would be needed if the decay channel were open for particular masses and lattice spacings\footnote{See \cite{Briceno:2017max} for a more detailed discussion of systematics.}.
\section{Lattice setup} \label{sec:lattice-setup}

We employ two $N_\mathrm{f}=2$ \acs{cls} ensembles \cite{Fritzsch:2012wq,Heitger:2013oaa}, characterized by the Wilson-plaquette gauge action and two mass-degenerate flavors of Wilson quarks with non-perturbative $\order{a}$ improvement. The main parameters appear in \cref{tab:cls-nf2-ensembles}. The charm-quark mass was fixed such that $m_{\PDs} = m_{\PDs,\text{phys}} = \qty{1968}{\mega\eV}$, while the gauge configurations were generated using a modified version of the \acs{ddhmc} algorithm \cite{Luscher:2003qa,Luscher:2004pav,Luscher:2007es}. We employ sequential fermion propagators with wall sources in a single time slice, diluted in time and spin. Besides, the scale is set using $f_{\PK}$ \cite{Fritzsch:2012wq}. In our simulations there are always one charm-quark and one charm-antiquark. We consider each of them to correspond to a different flavor, such that $\APcharm \Gamma \Pcharm^\prime$ is not a flavor singlet bilinear anymore. As a consequence, any Wick contraction where the charm-quarks annihilate is no longer possible, reducing the number of diagrams to compute.
\begin{table}
    \centering
    \begin{tabular}{cccc ccc}
        \toprule
        {id} & $\beta$ & $a~[\unit{\femto\meter}]$ & $L/a$ & $\kappa_{\ell}$ & $m_{\Ppi}~[\unit{\mega\eV}]$ & $m_{\Ppi}L$ \\
        \midrule
        D5 & 5.3 & 0.0658(7)(7) & 24 & 0.13625 & 449 & 3.6 \\
        E5 & 5.3 & 0.0658(7)(7) & 32 & 0.13625 & 437 & 4.7 \\
        \bottomrule
    \end{tabular}
    \caption[N\textsubscript{f}=2 ensembles]{The \ac{cls} ensembles used. From left to right, the ensemble label, the bare strong coupling, the lattice spacing as determined in \cite{Fritzsch:2012wq}, the number of lattice sites in every spatial direction ($T=2L$ in every ensemble), the value of $\kappa_{\ell}$, the approximate value of the pion mass \cite{DellaMorte:2017dyu}, and $m_{\Ppi}L$.}
    \label{tab:cls-nf2-ensembles}
\end{table}
The main requirement to extract the hadronic mixing is to make $\Pgya$ and $\APD\PD$ energy degenerate. To do that, we apply \acp{ptbc} \cite{Sachrajda:2004mi} to the charm quark. This transformation leaves the $\APcharm\Pcharm$ system at rest while we boost the $\PD$ mesons in opposite directions. Note that at the operator level, $\Pcharm$ and $\APcharm$ are related by $\APcharm=\Pcharm^\dagger \gamma_4$. Therefore, if we give momentum $\bar{p}$ to the quark, the anti-quark gets automatically momentum $-\bar{p}$, which means we cannot simulate configurations where both quark and antiquark have the same momentum. As a side effect, leaving the light-quarks with \acp{apbc} preserves isospin symmetry, and the final states $\APDzero\PDzero$ and $\PDminus\PDplus$ remain mass-degenerate in our simulations. We apply the same twist angle in every spatial direction, such that the $\PD$-meson 3-momentum is $a\bar{p} = a\theta/L \left(1,1,1\right)$. \Cref{tab:d-meson-spectrum-d5,tab:d-meson-spectrum-e5} show the probed values of the twist angle, and \cref{sec:analysis} gives a simple polynomial interpolation of our results, which can be used to find the decay width at any virtuality close to the on-shell condition.

We apply two levels of Gaussian smearing (20 and 50 iterations) to the charm-quark fields, but only one (50 iterations) to the light-quarks. Since the amplitudes of the various correlators must cancel exactly for the method developed in \cite{McNeile:2000xx} to work, the triangle diagram must have 50 iterations for all quarks at the source, where we locate the $\PD$ mesons, and either 20 or 50 at the sink, where the charmonium is located. The diagrams with exclusively $\PD$-mesons have only 50 iterations, and the charmonium two-point function has all possible combinations. We apply ten iterations of \acs{ape} smearing to the links employed in the Gaussian smearing to reduce short-distance fluctuations.

We use the following interpolators at zero momentum to create $\APcharm\Pcharm$ states with $\text{J}^{\text{PC}}=1^{--}$ quantum numbers,
\begin{equation}
    \label{eq:cc-operators}
    \begin{aligned}
        & O_1^\psi(\bar{0},t) = \sum_{\bar{x}} \APcharm_{\bar{x}} \gamma_i \Pcharm_{\bar{x}}, &
        & O_2^\psi(\bar{0},t) = \sum_{\bar{x}} \APcharm_{\bar{x}} \gamma_4 \gamma_i \Pcharm_{\bar{x}},
        \\
        & O_5^\psi(\bar{0},t) = \sum_{\bar{x}} \APcharm_{\bar{x}} \lva{\nabla}_i \gamma_i \va{\nabla}_i \Pcharm_{\bar{x}}, &
        & O_6^\psi(\bar{0},t) = \sum_{\bar{x}} \APcharm_{\bar{x}} \lva{\nabla}_i \gamma_4\gamma_i \va{\nabla}_i \Pcharm_{\bar{x}},
    \end{aligned}
\end{equation}
where $\bar{x}\in\mathbb{Z}^3$ is the position vector, and $O_1$, $O_2$, $O_5$, and $O_6$ label the operators following the structure from \cite{Mohler:2012na}. The aforementioned operators are built with the appropriate symmetries and quantum numbers to overlap with the desired physical states. However, each individual operator couples to an infinite tower of states with its quantum numbers, not only the ones we are interested in, and the coupling strength can vary. To solve these issues and build an operator that maximizes the coupling with $\Pgya$, we form a matrix with the correlators of the operators $O_i^\psi$ and solve the associated \ac{gevp} as described in \cref{sec:analysis}. This method creates orthogonal combinations of operators projected on each individual state of the tower. Note that how well the \ac{gevp} spectrum resembles the experiment depends on the diversity of the operators chosen. For example, we observe that the use of odd $T$ interpolators allows us to disentangle $\Pgyii$ and $\Pgya$ \cite{Mohler:2012na}. Looking at logical improvements, one can, in principle, enlarge the \ac{gevp} matrix and include several particle states like the operator $O^{\APD\PD}$ described in the following. However, the computational cost of the extra correlators quickly becomes too large. One last note is that, due to the reduction from Lorentz to hypercubic symmetry on the lattice, the operators also overlap with higher momentum states\footnote{See \cite{Mohler:2012na} for a comprehensive table of operators and their overlapping quantum numbers}. However, these are quite heavy and do not distort the lower spectrum.

\begin{figure}
    \centering
    \begin{subfigure}[t]{0.49\textwidth}
        \centering
        \includegraphics[scale=1]{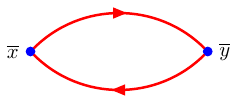}
        \caption{}
        \label{fig:propagator-cc-connected}
    \end{subfigure}
    \begin{subfigure}[t]{0.49\textwidth}
        \centering
        \includegraphics[scale=1]{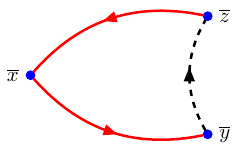}
        \caption{}
        \label{fig:decay-connected}
    \end{subfigure}
    \begin{subfigure}[t]{0.49\textwidth}
        \centering
        \includegraphics[scale=1]{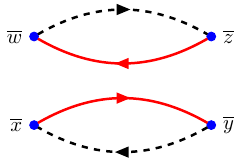}
        \caption{}
        \label{fig:propagator-dd-discconnected-free}
    \end{subfigure}
    \begin{subfigure}[t]{0.49\textwidth}
        \centering
        \includegraphics[scale=1]{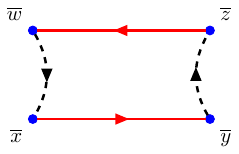}
        \caption{}
        \label{fig:propagator-dd-connected-u-annihilation}
    \end{subfigure}
    \caption[Wick contractions needed]{Main Wick contractions in this project. \Cref{fig:propagator-cc-connected} is the heavy meson propagator, \cref{fig:decay-connected} the triangle diagram for the hadronic mixing, \cref{fig:propagator-dd-discconnected-free} the direct $\APD\PD$ propagator, and \cref{fig:propagator-dd-connected-u-annihilation} the box $\APD\PD$ propagator. Solid lines indicate a charm propagator, dashed lines a light quark, and blue dots operator insertions in position space. Diagrams with charm annihilation are not considered.}
    \label{fig:propagators}
\end{figure}

To interpolate the $\APD\PD$ system on the $p$-wave configuration with isospin symmetry, we employ
\begin{equation}
    O^{\APD\PD}(\bar{p},t)
    =
    \APDzero(\bar{p},t)\PDzero(-\bar{p},t) - \APDzero(-\bar{p},t)\PDzero(\bar{p},t),
\end{equation}
where each $\PD$-meson is given by
\begin{equation}
    \APDzero(\bar{p},t) = \sum_{\bar{x}} e^{-\iu a\bar{x}\bar{p}} \APcharm_{\bar{x}} \gamma_5 \Pup_{\bar{x}}
    \quad \text{and} \quad
    \PDzero(\bar{p},t)  = \sum_{\bar{x}} e^{-\iu a\bar{x}\bar{p}} \APup_{\bar{x}} \gamma_5 \Pcharm_{\bar{x}}.
\end{equation}
The momentum $\bar{p}\in\mathbb{R}^3$ is chosen arbitrarily thanks to \acp{ptbc}. A similar discussion to that of charmonium applies here. However, the $\APD\PD$ spectrum is expected to have much larger gaps such that excited states have only a small effect on the extracted levels. That is why we consider a single interpolator.
Next, we turn to the correlators, which appear in \cref{fig:propagators}. Consider light and heavy propagators $U_{\bar{x}\bar{y}}$ and $H_{\bar{x}\bar{y}}$, respectively, between 3-vector positions $\bar{x}$ and $\bar{y}$. To compute the spectrum of charmonium on the lattice, we require a symmetric matrix
\begin{equation}
  P^\psi_{ij}(t) = \frac{\text{sign}}{V} \sum_{\bar{x},\bar{y}} \expval*{\APcharm_{\bar{x}} \Gamma_i \Pcharm_{\bar{x}} \APcharm_{\bar{y}} \Gamma_j \Pcharm_{\bar{y}}}
  = - \frac{\text{sign}}{V} \sum_{\bar{x},\bar{y}} \expval*{\tr(H_{\bar{x}\bar{y}}\Gamma_jH_{\bar{y}\bar{x}}\Gamma_i)}_\text{G}.
\end{equation}
The sign depends on the $\Gamma$ structure at the source, $\text{sign}=+1$ if $\Gamma_j=\gamma_j$, and $\text{sign}=-1$ if $\Gamma_j=\gamma_0\gamma_j$. This correlator corresponds to \cref{fig:propagator-cc-connected}. We also use one real $\PD$-meson correlator,
\begin{equation}
  P^{\PD}(\bar{p},t) = -\frac{1}{V} \sum_{\bar{x},\bar{y}} \cos[a\bar{p}(\bar{x}-\bar{y})] \expval*{\tr(U_{\bar{x}\bar{y}}\Gamma H_{\bar{y}\bar{x}} \Gamma)}_\text{G},
\end{equation}
which can be obtained replacing a heavy-quark by a light-quark propagator in \cref{fig:propagator-cc-connected}. The transition matrix element, shown in \cref{fig:decay-connected}, is a purely imaginary number,
\begin{equation}
    T_i(\bar{p},t) = \frac{2\iu}{V} \sum_{\bar{x},\bar{y},\bar{z}} \sin[a\bar{p}(\bar{y}-\bar{z})] \expval*{\tr(\Gamma H_{\bar{x}\bar{y}} \Gamma U_{\bar{y
    }\bar{z}} \Gamma H_{\bar{z}\bar{x}}) }_\text{G}.
\end{equation}
We use \ac{ptbc} to fix the $\PD$-meson momentum in the previous correlation functions. Finally, the $\APD\PD$ propagator can be divided in two pieces,
\begin{multline}
    P^{\PD\PD}(\bar{p},t)
    = \frac{1}{V} \sum_{\bar{w},\bar{x},\bar{y},\bar{z}} \sin[a\bar{p}(\bar{w}-\bar{x})] \sin[a\bar{p}(\bar{y}-\bar{z})]
    \\
    \Big(
        -2\expval*{\tr(H_{\bar{x}\bar{y}} \Gamma U_{\bar{y}\bar{x}} \Gamma) \tr(H_{\bar{z}\bar{w}} \Gamma U_{\bar{w}\bar{z}} \Gamma)}_\text{G}
        \\
        +4\expval*{\tr(H_{\bar{x}\bar{y}} \Gamma U_{\bar{y}\bar{z}} \Gamma H_{\bar{z}\bar{w}} \Gamma U_{\bar{w}\bar{x}} \Gamma)}_\text{G}
    \Big),
\end{multline}
where the first term is the direct contribution in \cref{fig:propagator-dd-discconnected-free}, and the second term represents the box component in \cref{fig:propagator-dd-connected-u-annihilation}. We cannot compute the full $\HepProcess{\APD\PD\to\APD\PD}$ correlator employing \acp{ptbc}, unfortunately. It is possible to see this writing the trigonometric functions as phases, and associating each one with its corresponding heavy quark. The configuration of momentum for some contributions to the direct and box diagram are incompatible with \acp{ptbc}. To bypass this problem, we split the correlator in two different pieces. The main component, 
\begin{multline}
    P^{\PD\PD}_\text{u}(\bar{p},t)
    =
    \frac{1}{V} \expval{\abs{\expval*{\APDzero(-\bar{p},t) \PDzero(\bar{p},0)}_\text{F}}^2}_\text{G}
    \\
    + \frac{2}{V} \expval{\APDzero(-\bar{p},t)\PDzero(\bar{p},t)\APDzero(-\bar{p},0)\PDzero(\bar{p},0)}_\text{box},
\end{multline}
which can be computed using \acp{ptbc}, and another piece, that we compute using Fourier modes (note the different momentum signs),
\begin{multline}
    P^{\PD\PD}_\text{f}(\bar{p},t)
    =
    \frac{1}{V} \expval{\abs{\expval*{\APDzero(-\bar{p},t) \PDzero(-\bar{p},0)}_\text{F}}^2}_\text{G}
    \\
    + \frac{2}{V} \expval{\APDzero(-\bar{p},t)\PDzero(\bar{p},t)\APDzero(\bar{p},0)\PDzero(-\bar{p},0)}_\text{box}.
\end{multline}
To compute the full $\APD\PD$ propagator at a given momentum $p$, we interpolate linearly $P^{\PD\PD}_\text{f}$ using its value at rest and on the first Fourier mode. Then,
\begin{equation}
    P^{\PD\PD}(\bar{p},t) = P^{\PD\PD}_{\text{u}}(\bar{p},t) - P^{\PD\PD}_{\text{f}}(\bar{p},t).
\end{equation}
At rest, both components exactly cancel each other, but as one increases the momentum, the latter becomes quickly subdominant.
\section{Analysis} \label{sec:analysis}

In this section, we present our analysis of the $N_f=2$ dataset detailed in \cref{sec:lattice-setup}\footnote{Regarding the statistical analysis, we employ the package \textit{pyerrors} \cite{Joswig:2022qfe,Ramos:2018vgu,Schaefer:2010hu}, which estimates the auto-correlation matrix as explained in \cite{Wolff:2003sm}.}.
\begin{figure}
    \centering
    \begin{subfigure}[t]{0.49\textwidth}
        \centering
        \includegraphics[scale=1]{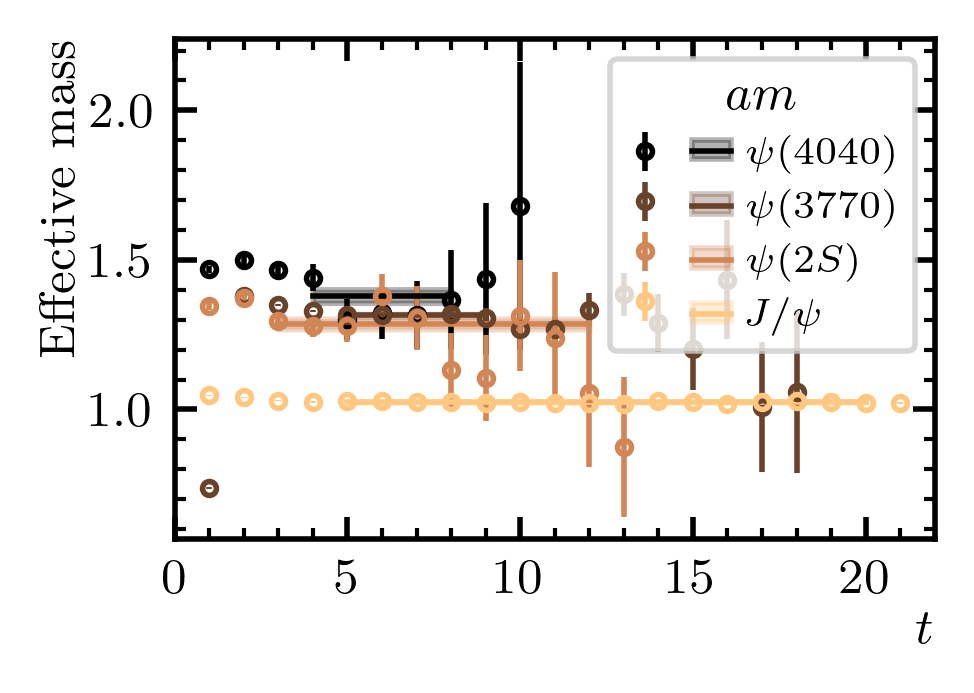}
    \end{subfigure}
    \begin{subfigure}[t]{0.49\textwidth}
        \centering
        \includegraphics[scale=1]{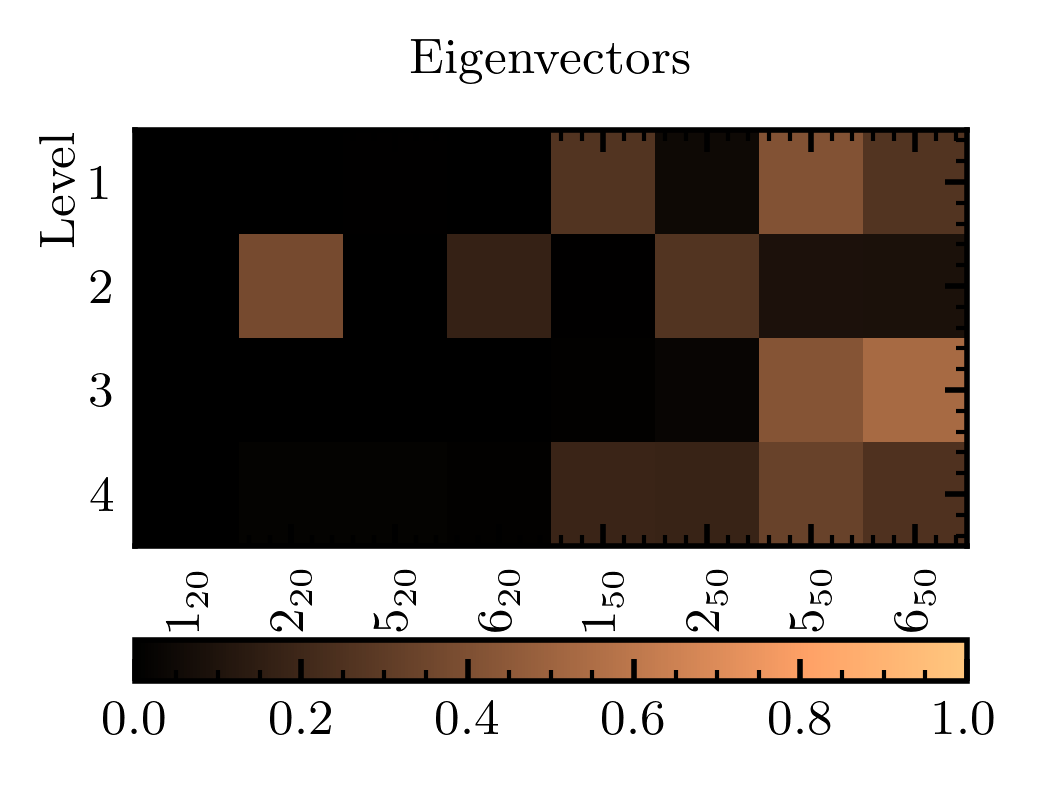}
    \end{subfigure}
    \caption[\ac{gevp} eigenvalues and eigenvectors]{
    The charmonium spectrum (LEFT) and eigenvectors (RIGHT) as obtained from an $8 \times 8$ \ac{gevp} on ensemble D5. The labels refer to the interpolator (see \cref{eq:cc-operators}) and the number of smearing iterations.}
    \label{fig:cc-gevp}
\end{figure}
In experiment and in lattice simulations with all dynamical flavors, one obtains a spectrum that includes the mixing between different particles.
However, in our case, the transition $\HepProcess{\Pgya\to\APD\PD}$, with the final pair of mesons in a p-wave configuration, is kinematically forbidden because the first Fourier mode is much heavier than the states considered. This allows us to obtain an unmixed spectrum and compute the hadronic transition directly, making the comparison to phenomenological models easier.
By adjusting the charm-quark twist-angle $\theta$, we set the final state energy $E_{\APD\PD}$. We assume that the $\APD\PD$ mesons do not interact with each other, so that $E_{\PD\APD} = 2E_{\PD}$, and that the relativistic dispersion in the continuum holds at finite lattice spacing,
\begin{equation}
  E_{\PD}^2=m_{\PD}^2+3\left(\dfrac{\theta}{L}\right)^2.
\end{equation}
Note that we set \acp{ptbc} for each space direction, creating an isotropic momentum $\bar{p} = (1,1,1)~\theta/L$.
The threshold condition in the \acs{cm} frame reads $m_{\Pgya}=2E_{\PD}$, and the corresponding twist angle is given by
\begin{equation}
    \label{eq:theta-vs-mass}
    \dfrac{\theta_0}{L}=\sqrt{\dfrac{1}{12}\left[m_{\psi}^2-(2m_{\PD})^2\right]}.
\end{equation}
The next step is to determine $am_{\psi}$ and $am_{\PD}$ in the ensembles D5 and E5. First, we form an 8x8 \ac{gevp} with the bilinears given in \cref{sec:lattice-setup} and two levels of Gaussian smearing with 20 and 50 iterations. We solve the linear system for the first four vector states in time slices $t_0/a=4$ and $t/a=6$, fulfilling the rule $2t_0 \geq t$ necessary to minimize systematic uncertainties \cite{Blossier:2009kd}.
In this regime, the main systematic error in the extraction of the energy levels vanishes like $\exp(-t[E_{N+1}-E_n])$ \cite{Blossier:2009kd}, where $E_n$ is the energy level computed and $E_{N+1}$ is the first state not included in the \ac{gevp}. In our case, $N=8$ but we only show results $n \leq 4$ because the signal is lost too fast.
The eigenvectors of each energy level appear in \cref{fig:cc-gevp} normalized to $(v_n)_i(v_n)_i/(v_n^{\tmat} v_n)$.
The index $i$ runs over the vector entries, and it is not summed. We observe that larger smearing interpolates our states better, and we need odd $T$ interpolators to separate $\Pgyii$ from $\Pgya$.
\begin{figure}
	\centering
  	\begin{subfigure}[t]{0.49\textwidth}
  		\centering
  		\includegraphics[scale=1]{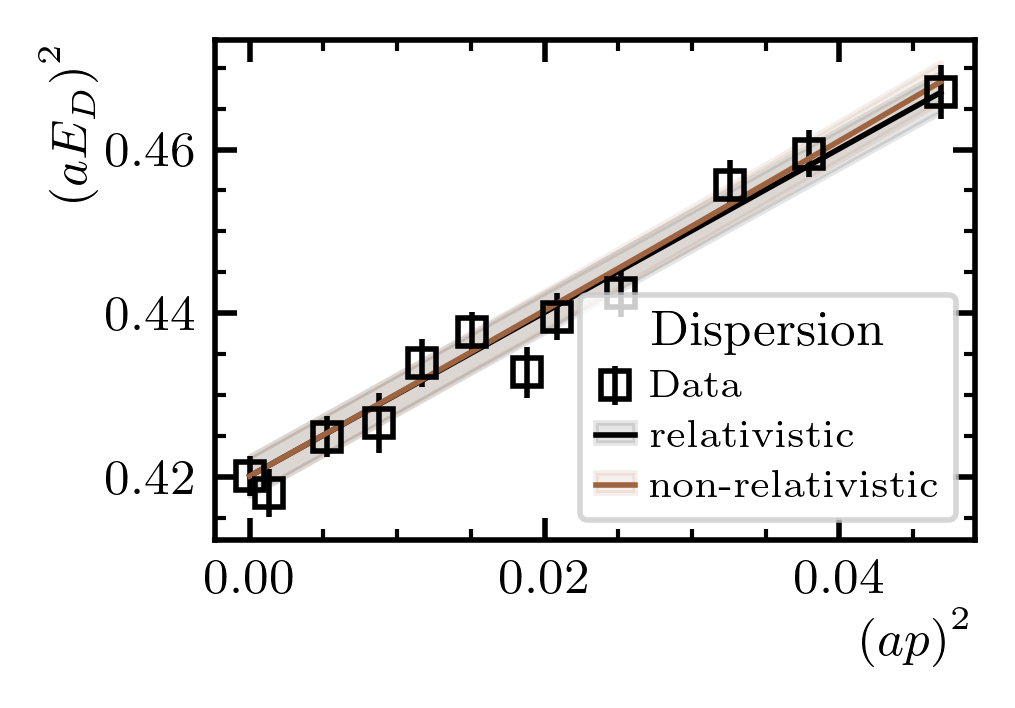}
  	\end{subfigure}
  	\begin{subfigure}[t]{0.49\textwidth}
  		\centering
  		\includegraphics[scale=1]{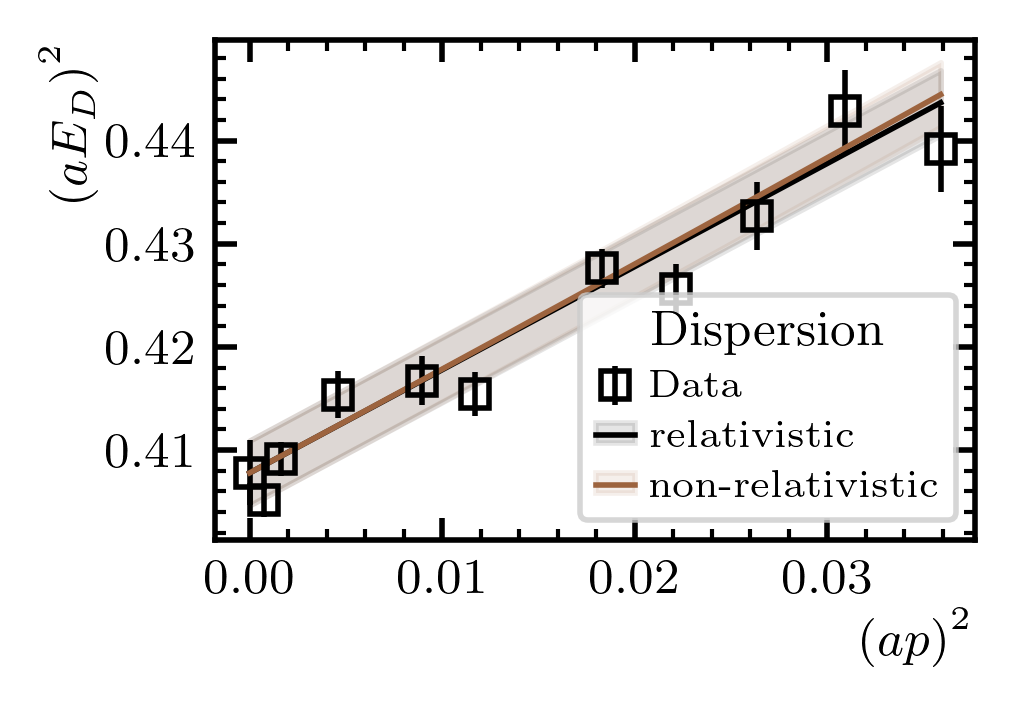}
  	\end{subfigure}
  \caption[D-meson spectrum]{The $\PD$-meson energy as a function of the twist angle compared to the dispersion relation in the continuum on ensembles D5 (LEFT) and E5 (RIGHT). We observe the data can be described well by the non-relativistic dispersion relation including the rest mass.}
  \label{fig:dd-spectrum}
\end{figure}
Once we have obtained the eigenvectors, we project $C(t)$ to a particular level to obtain the eigenvalues, which show a residual time dependence; see \cref{fig:cc-gevp}. To remove the latter, we compute the effective mass of the eigenvalue and fit it to a constant. The resulting spectrum can be seen in \cref{tab:cc-spectrum} in lattice and physical units.
\begin{table}
  	\centering
	\begin{tabular}{ll ll l}
		\toprule
		Level & Fit Interval & $\chi^2/\text{dof}$ & $am_{\psi}$ & $m_{\psi}~[\unit{\mega\eV}]$ \\
		\midrule
		\begin{filecontents*}{tables/D5-charmonium-spectrum.csv}
,Level,FitIntervalA,FitIntervalB,Chisquare,Dof,ChisquareByDof,Pvalue,EnergyLattice,EnergyMeV
1,1,5,20,25.96,15,1.73,0.04,1.02423(63),3074.1(2.4)
2,2,3,12,5.93,9,0.66,0.75,1.286(17),3859(52)
3,3,5,9,0.34,4,0.08,0.99,1.3148(43),3946(13)
4,4,4,8,3.88,4,0.97,0.42,1.379(22),4139(67)
\end{filecontents*}
\csvreader[head to column names, late after line=\\]{tables/D5-charmonium-spectrum.csv}{}
		{\Level & [\FitIntervalA, \FitIntervalB] & \Chisquare/\Dof & \EnergyLattice & \EnergyMeV}
		\bottomrule
    \end{tabular}
	\caption[Charmonium spectrum on D5]{Charmonium spectrum on ensemble D5. We employ the same values on E5.}
	\label{tab:cc-spectrum}
\end{table}
Next, we determine the $\PD$-meson energy for various twist angles $\theta$ using a single-exponential fit to the correlators of the $\APcharm\gamma_5\Pup$ interpolator. The fit results appear in \cref{tab:d-meson-spectrum-d5,tab:d-meson-spectrum-e5}, and we compare the energies with their continuum dispersion relation in \cref{fig:dd-spectrum}, which shows that $E_{\PD}$ is free of lattice artifacts and follows the non-relativistic dispersion relation. This proves our initial assumption for the momentum dependence of $E_{\PD}$.
\begin{table}
	\centering
	\begin{tabular}{ll ll ll l}
		\toprule
		$\theta$ & $\sqrt{3} \theta/L$ & Fit Interval & $\chi^2/\text{dof}$ & $\abs{a^3 A_{\PD}}^2 \times 10^7$ & $aE_{\PD}$ & $E_{\PD}~[\unit{\mega\eV}]$ \\
		\midrule
		\begin{filecontents*}{tables/D5-dmeson-spectrum.csv}
,Theta,Momentum,FitIntervalA,FitIntervalB,Chisquare,Dof,ChisquareByDof,AmplitudeSeven,EnergyLattice,EnergyMeV
0,0.0,0.0,7,15,1.04,7,0.15,2708(61),0.6482(18),1945.3(5.6)
1,0.5,0.036,7,20,12.29,12,1.02,2707(64),0.6466(23),1940.5(6.9)
2,1.0,0.072,7,18,8.16,10,0.82,2737(58),0.6519(19),1956.5(5.9)
3,1.3,0.094,7,18,8.67,10,0.87,2642(63),0.6531(28),1960.3(8.4)
4,1.5,0.108,5,20,19.35,14,1.38,2742(54),0.6588(22),1977.2(6.8)
5,1.7,0.123,7,19,12.74,11,1.16,2692(65),0.6615(19),1985.5(5.9)
6,1.9,0.137,7,19,5.07,11,0.46,2559(63),0.6579(24),1974.5(7.2)
7,2.0,0.144,7,20,10.2,12,0.85,2743(66),0.6630(22),1989.9(6.6)
8,2.2,0.159,7,20,18.67,12,1.56,2567(70),0.6652(22),1996.5(6.7)
9,2.5,0.18,7,20,16.34,12,1.36,2494(60),0.6750(22),2025.9(6.8)
10,2.7,0.195,7,18,21.58,10,2.16,2647(67),0.6778(21),2034.4(6.4)
11,3.0,0.217,6,20,11.24,13,0.86,2488(53),0.6834(24),2051.2(7.3)
\end{filecontents*}
\csvreader[head to column names, late after line=\\]{tables/D5-dmeson-spectrum.csv}{}
		{\Theta & \Momentum & [\FitIntervalA, \FitIntervalB] & \Chisquare/\Dof & \AmplitudeSeven & \EnergyLattice & \EnergyMeV}
		\bottomrule
\end{tabular}
  \caption[D-meson spectrum on D5]{$\PD$-meson spectrum on D5. From left to right, the twist angle and momentum, fit interval, fit quality, amplitude of the exponential, $aA_{\PD}$, and energy of the $\PD$-meson.}
  \label{tab:d-meson-spectrum-d5}
\end{table}
\begin{table}
	\centering
	\begin{tabular}{ll ll ll l}
		\toprule
		$\theta$ & $\sqrt{3} \theta/L$ & Fit Interval & $\chi^2/\text{dof}$ & $\abs{a^3 A_{\PD}}^2 \times 10^7$ & $aE_{\PD}$ & $E_{\PD}~[\unit{\mega\eV}]$ \\
		\midrule
		\begin{filecontents*}{tables/E5-dmeson-spectrum.csv}
,Theta,Momentum,FitIntervalA,FitIntervalB,Chisquare,Dof,ChisquareByDof,AmplitudeSeven,EnergyLattice,EnergyMeV
0,0.0,0.0,17,22,3.09,4,0.77,2565(109),0.6386(25),1914.9(7.4)
1,0.5,0.027,11,26,15.65,14,1.12,2525(44),0.6365(13),1908.8(4.0)
2,0.75,0.041,11,26,10.33,14,0.74,2592(45),0.6396(13),1918.1(4.0)
3,1.25,0.068,11,26,2.77,14,0.2,2604(57),0.6445(18),1932.6(5.4)
4,1.75,0.095,12,22,10.1,9,1.12,2465(69),0.6455(18),1935.8(5.6)
5,2.0,0.108,12,22,15.89,9,1.77,2462(65),0.6445(17),1932.8(5.1)
6,2.5,0.135,10,20,4.29,9,0.48,2512(50),0.6539(15),1961.0(4.5)
7,2.75,0.149,12,25,26.67,12,2.22,2334(60),0.6524(18),1956.4(5.6)
8,3.0,0.162,15,20,3.32,4,0.83,2477(97),0.6578(25),1972.5(7.6)
9,3.25,0.176,15,20,2.01,4,0.5,2588(119),0.6655(30),1995.6(9.1)
10,3.5,0.189,15,20,2.84,4,0.71,2292(114),0.6627(31),1987.3(9.5)
\end{filecontents*}
\csvreader[head to column names, late after line=\\]{tables/E5-dmeson-spectrum.csv}{}
		{\Theta & \Momentum & [\FitIntervalA, \FitIntervalB] & \Chisquare/\Dof & \AmplitudeSeven & \EnergyLattice & \EnergyMeV}
		\bottomrule
\end{tabular}
  \caption[D-meson spectrum on E5]{$\PD$-meson spectrum on E5. Format as in \cref{tab:d-meson-spectrum-d5}.}
  \label{tab:d-meson-spectrum-e5}
\end{table}
Unlike in the charmonium spectrum, the light-quarks play a more significant role in the determination of the $\PD$ mass, as well as finite-size effects and the missing continuum extrapolation. Plugging the values of $am_\psi$ and $am_{\PD}$ into \cref{eq:theta-vs-mass}, we obtain the twist angle $\theta_0$ for the on-shell condition. The values appear in \cref{tab:results}.

\begin{figure}
	\centering
	\begin{subfigure}[t]{0.49\textwidth}
		\centering
		\includegraphics[scale=1]{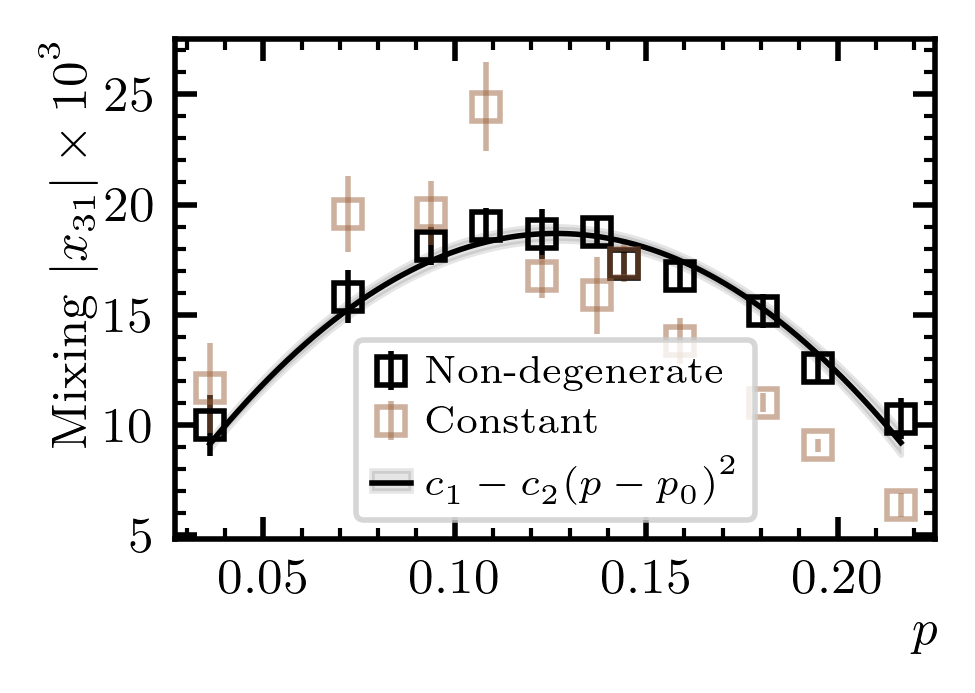}
	\end{subfigure}
	\begin{subfigure}[t]{0.49\textwidth}
		\centering
		\includegraphics[scale=1]{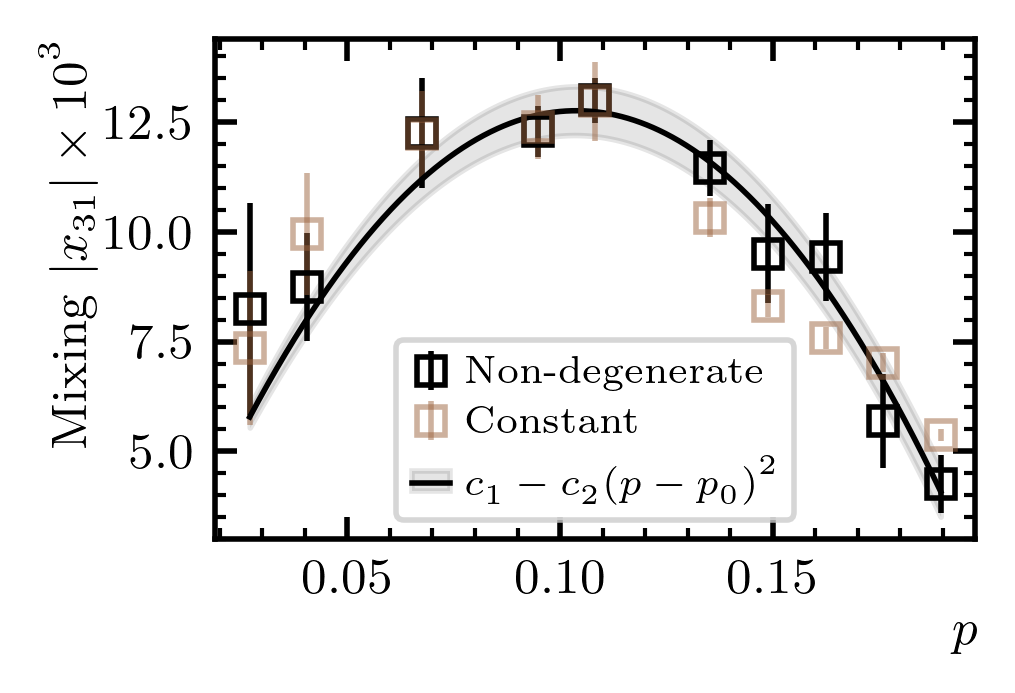}
	\end{subfigure}
	\caption[Mixing vs twist angle]{The mixing $\abs{x_{31}} \times 10^3$ for ensembles D5 (LEFT) and E5 (RIGHT). The dark points are obtained from a linear fit to the non-degenerate formula in \cref{eq:off-shell-ratio}, while the lighter points are fitted to a constant using \cref{eq:xt}. The band fits only the dark points.}
	\label{fig:mixing-vs-twist}
\end{figure}

If the on-shell condition is satisfied, we can apply \cref{eq:on-shell-ratio} to extract the mixing $x_{31}$. However, we can relax this condition and use \cref{eq:off-shell-ratio}. Both are linear fits and can be solved algebraically. We decide to fit the lattice data to the latter expression, including correlations. The fit results for ensembles D5 and E5 appear in \cref{tab:non-degenerate-d5,tab:non-degenerate-e5}, respectively.
There, we gather the fit interval in lattice units, the fit quality, as well as the values of the hadronic mixing and the constant $A$. The fit is plotted in \cref{fig:fit-slope}, where the ratio $\mathcal{R}$ appears as a function of time, and we indicate the fit range and the uncertainty.
\begin{table}
	\centering
	\begin{tabular}{l l l l l}
		\toprule
		$\theta$ & Fit Interval & $\sqrt{3}\theta/L$ & $a\abs{x_{31}}$ & $A$ \\
		\midrule
		\begin{filecontents*}{tables/D5-non-degenerate.csv}
,Chisquare,Dof,ChisquareByDof,Theta,Momentum,Mixing,Offset,TimeIntervalA,TimeIntervalB
0,27.23,30,0.91,0.5,0.036,0.0100(14),0.0010(29),2,5
1,27.23,30,0.91,1.0,0.072,0.0158(12),-0.0002(23),2,5
2,27.23,30,0.91,1.3,0.094,0.01811(87),-0.0028(17),2,5
3,27.23,30,0.91,1.5,0.108,0.01903(80),-0.0039(15),2,6
4,27.23,30,0.91,1.7,0.123,0.0187(11),-0.0044(21),2,7
5,27.23,30,0.91,1.9,0.137,0.01874(72),-0.0055(16),2,7
6,27.23,30,0.91,2.0,0.144,0.01731(73),-0.0050(19),2,7
7,27.23,30,0.91,2.2,0.159,0.01675(73),-0.0055(15),2,6
8,27.23,30,0.91,2.5,0.18,0.01516(77),-0.0066(16),2,5
9,27.23,30,0.91,2.7,0.195,0.01260(65),-0.0054(15),2,5
10,27.23,30,0.91,3.0,0.217,0.01029(92),-0.0070(16),2,5
\end{filecontents*}
\csvreader[head to column names, late after line=\\]{tables/D5-non-degenerate.csv}{}
		{\Theta & [\TimeIntervalA, \TimeIntervalB] & \Momentum & \Mixing & $\Offset$}
		\bottomrule
    \end{tabular}
	\caption[Hadronic mixing on ensemble D5]{Fit to the off-shell \cref{eq:off-shell-ratio} on ensemble D5. The constant $A$ is dimensionless. All data points are fitted together, giving $\chi^2/\text{dof} = 27.23 / 30$. A plot of these points appear in \cref{fig:mixing-vs-twist}.}
	\label{tab:non-degenerate-d5}
\end{table}
\begin{table}
	\centering
	\begin{tabular}{l l l l l}
		\toprule
		$\theta$ & Fit interval & $\sqrt{3}\theta/L$ & $a\abs{x_{31}}$ & $A$ \\
		\midrule
		\begin{filecontents*}{tables/E5-non-degenerate.csv}
,Chisquare,Dof,ChisquareByDof,Theta,Momentum,Mixing,Offset,TimeIntervalA,TimeIntervalB
0,34.23,27,1.27,0.5,0.027,0.0082(24),-0.013(15),6,10
1,34.23,27,1.27,0.75,0.041,0.0088(12),-0.0096(59),5,9
2,34.23,27,1.27,1.25,0.068,0.0122(13),-0.0152(79),6,9
3,34.23,27,1.27,1.75,0.095,0.01229(58),-0.0096(24),4,8
4,34.23,27,1.27,2.0,0.108,0.01300(51),-0.0122(24),4,8
5,34.23,27,1.27,2.5,0.135,0.01146(64),-0.0106(33),5,9
6,34.23,27,1.27,2.75,0.149,0.0095(11),-0.0003(72),6,10
7,34.23,27,1.27,3.0,0.162,0.0094(10),-0.0066(61),6,9
8,34.23,27,1.27,3.25,0.176,0.0057(11),0.0140(63),6,9
9,34.23,27,1.27,3.5,0.189,0.00426(67),0.0107(33),5,9
\end{filecontents*}
\csvreader[head to column names, late after line=\\]{tables/E5-non-degenerate.csv}{}
		{\Theta & [\TimeIntervalA, \TimeIntervalB] & \Momentum & \Mixing & $\Offset$}
		\bottomrule
    \end{tabular}
	\caption[Hadronic mixing on ensemble E5]{Fit to the off-shell \cref{eq:off-shell-ratio} on ensemble E5. The constant $A$ is dimensionless. All data points are fitted together, giving $\chi^2/\text{dof} = 34.23 / 27$. A plot of these points appear in \cref{fig:mixing-vs-twist}.}
	\label{tab:non-degenerate-e5}
\end{table}
As a next step, we plot the results of \cref{tab:non-degenerate-d5,tab:non-degenerate-e5} as a function of the $\PD$-meson momentum in \cref{fig:mixing-vs-twist}. The general structure of the plots was expected. From a maximum close to the on-shell condition, the matrix element quickly decreases as the momentum is varied. We also plot the fit to a constant from \cref{eq:xt}, which gives values in the same range, although we believe it is inferior to \cref{eq:off-shell-ratio} for the reasons we mentioned in \cref{sec:methodology}. We observe that the momentum dependence can be fitted by a parabola that vanishes at $p=0$,
\begin{equation}
	\label{eq:poly-model}
	\abs{x_{31}}(p) = c_1 - c_2 (p-p_0)^2,
\end{equation}
where $p = \sqrt{3}\theta/L$, and $p_0=\sqrt{c_1/c_2}$. We perform a correlated fit of the data points and obtain $ac_1 = \num{0.01869(32)}$, $c_2/a = \num{1.169(44)}$ with $\chi^2/\text{dof} = 34.39 / 39$ in ensemble D5, $ac_1 = \num{0.01276(56)}$, $c_2/a=\num{1.183(52)}$ with $\chi^2/\text{dof}=42.51 / 35$ on E5.

The choice of fit intervals to extract $a\abs{x_{31}}$ is paramount to obtain a good $\chi^2$ per degree of freedom. Taking into account all the twist angles, there are billions of options for reasonable fit intervals. We obtain 10000 samples $P[c_1]$, $P[c_2]$, and $P[\chi^2]$ for variables $c_1$, $c_2$, and $\chi^2$, respectively, choosing random fit intervals for each twist angle in \cref{eq:off-shell-ratio}. We use these samples as a guide to choose the final fit intervals shown in \cref{tab:non-degenerate-d5,tab:non-degenerate-e5}. The $\chi^2$ and degrees of freedom shown under \cref{eq:poly-model} are the sum of \cref{eq:on-shell-ratio,eq:poly-model} fits.

With \cref{eq:poly-model}, we can find the decay width of $\Pgya$ employing \cref{eq:decay-width-lattice} directly.
Furthermore, the energy shift of the spectrum can be determined using perturbation theory. For the on-shell condition, the states are separated by $E \to E \pm \abs{x_{31}}$. For a generic transition, we use \cref{eq:transfer-matrix}
\footnote{Although the only object defined for off-shell kinematics is the amplitude $\abs{x_{31}}$, this is not a physical observable, and different approaches may normalize this quantity in a different way ---for instance, the ${}^3P_0$ quark-model described in \cref{sec:quark-model} has different units than our calculation. Then, to compare the entire kinematic range with other determinations, we remove the condition of energy conservation from the decay width, in such a way as to compare objects with the same units. Obviously, this is purely a theoretical comparison between two methods, and the decay width is only properly defined on-shell.}.
The values on-shell appear in \cref{tab:results}, which summarizes our main results. First, we observe that the twist angle $\theta$ for the on-shell condition varies a factor two between D5 and E5. This is due to the small difference between the spectrum of the two ensembles, which can be traced back to \acp{fve}. Therefore, an important conclusion is that \acp{fve} change the on-shell condition quite dramatically. Associated with the twist angle, the $\PD$-meson has a momentum that, although large, still allows for a non-relativistic analysis. Next, the hadronic mixing $a\abs{x_{31}}$ directly gives the energy shift for the spectrum. The heavier state becomes heavier by an amount $+\epsilon$ and the lighter system becomes lighter by the same amount, at first order in perturbation theory. Then, we indicate the decay width obtained on the lattice. The results for both ensembles agree very well, and we expect the difference to come from \acp{fve}, which at our level of precision are still not very relevant. From the two ensembles, E5 is larger and its results should be closer to infinite volume. Therefore, we take the value of $\Gamma~[\unit{\mega\eV}]$ in E5 shown in \cref{tab:results} as our best estimate of the decay width of $\Pgya$. This should be compared to the experimental result $\Gamma(\HepProcess{\Pgya\to\APD\PD}) = \qty{27.2(1.0)}{\mega\eV}$ \cite{ParticleDataGroup:2022pth}.
Of course, one should bear in mind that ours is an exploratory study, utilizing only two ensembles with non-physical pion masses, which can affect the $\PD$-meson spectrum, and we have not taken the continuum limit either. However, our results demonstrate that it is possible to extract the decay width of a charmonium excitation using the method developed in \cite{McNeile:2000xx}.

\begin{figure}
	\centering
	\begin{subfigure}[t]{0.49\textwidth}
		\centering
		\includegraphics[scale=1]{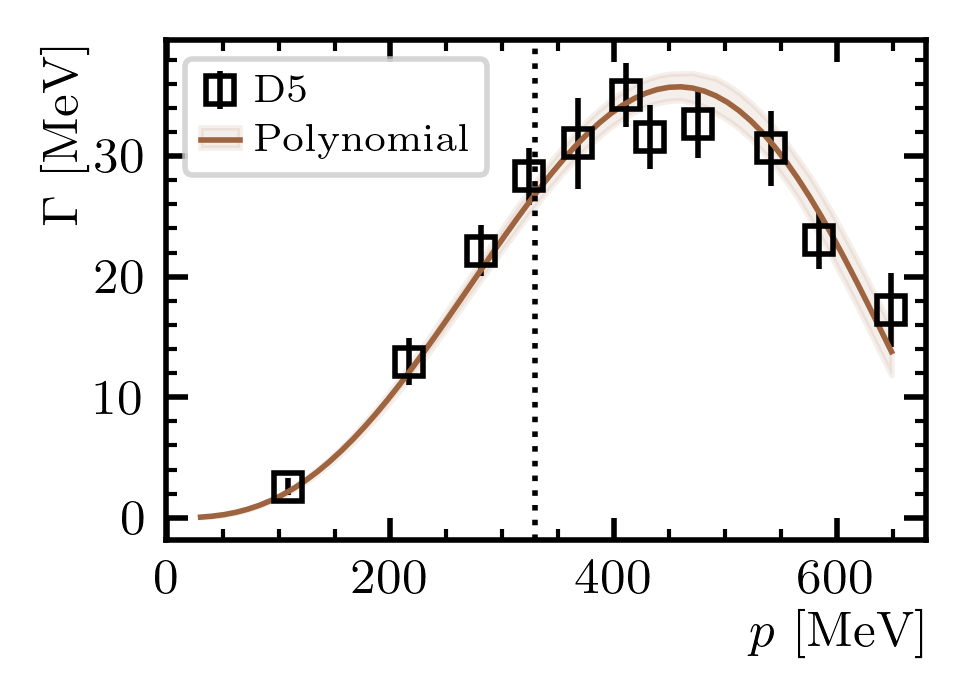}
	\end{subfigure}
	\begin{subfigure}[t]{0.49\textwidth}
		\centering
		\includegraphics[scale=1]{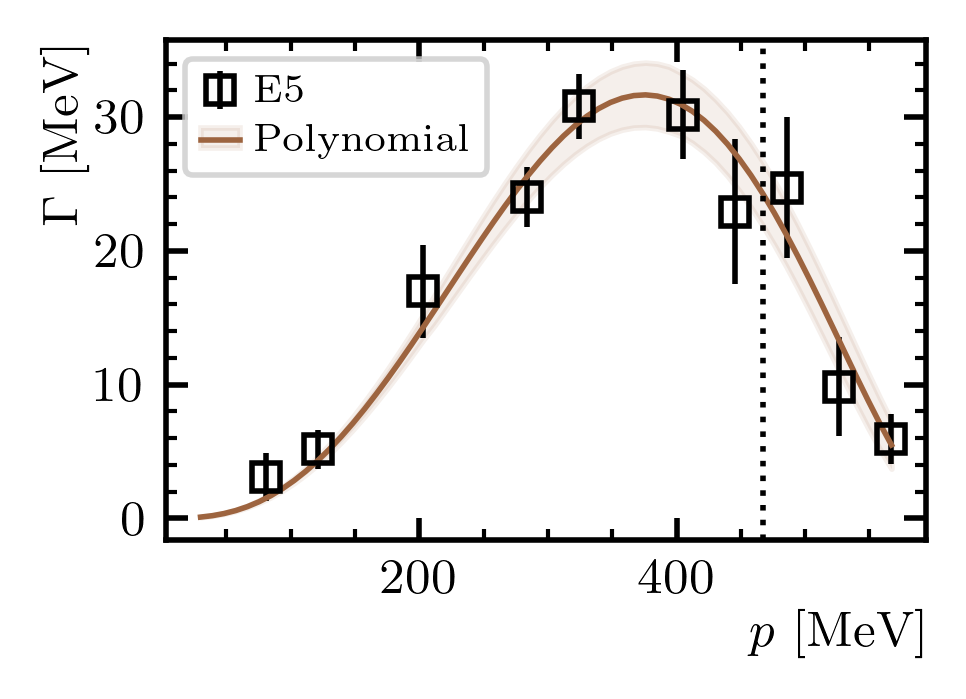}
	\end{subfigure}
	\caption[Decay width]{Decay width on ensembles D5 (LEFT) and E5 (RIGHT) as a function of the $\PD$-meson momentum. We fit the lattice data to \cref{eq:poly-model,eq:decay-width-lattice} (contiguous line). The vertical line marks the on-shell condition.}
	\label{fig:decay-width}
\end{figure}

Lastly, one can qualitatively explain the lattice data using some analytic method. We resort to the ${}^3P_0$ quark model described in \cref{sec:quark-model}. Comparing \cref{fig:test_quark_model,fig:decay-width}, we observe that the quark model reproduces the main features of our data, mainly its bell shape as a function of the off-shell $\PD$-meson momentum. In this model, one can play with the coupling $\beta$, that varies the height of the curve, and the parameters of the harmonic oscillators. Although it provides us with a physical understanding of the process, the model cannot provide a good description in terms of fit quality.
\begin{table}
	\centering
	\begin{tabular}{lll lll l}
		\toprule
		Ensemble & $\theta_0~[\unit{\radian}]$ & $p~[\unit{\mega\eV}]$ & $a\abs{x_{31}}$ & $\epsilon~[\unit{\mega\eV}]$ & $\Gamma/a$ & $\Gamma~[\unit{\mega\eV}]$ \\
		\midrule
		\begin{filecontents*}{tables/D5-on-shell_decay.csv}
,Ensemble,Momentum,MomentumMeV,Theta,Mixing,Shift,ShiftMeV,DecayWidth,DecayWidthMeV
0,D5,0.110(17),329(52),1.52(24),0.01837(73),0.01837(73),55.1(2.2),0.0089(21),26.8(6.2)
\end{filecontents*}
\csvreader[head to column names, late after line=\\]{tables/D5-on-shell_decay.csv}{}{D5 & \Theta & \MomentumMeV & \Mixing & \ShiftMeV & \DecayWidth & \DecayWidthMeV}%
		\begin{filecontents*}{tables/E5-on-shell_decay.csv}
,Ensemble,Momentum,MomentumMeV,Theta,Mixing,Shift,ShiftMeV,DecayWidth,DecayWidthMeV
0,E5,0.156(13),468(40),2.89(25),0.0095(16),0.0095(16),28.5(4.9),0.0081(21),24.2(6.4)
\end{filecontents*}
\csvreader[head to column names, late after line=\\]{tables/E5-on-shell_decay.csv}{}{E5 & \Theta & \MomentumMeV & \Mixing & \ShiftMeV & \DecayWidth & \DecayWidthMeV}
		\bottomrule
	\end{tabular}
	\caption[On-shell results]{On-shell results. The twist angle $\theta_0$ where the energy is conserved, the associated $\PD$-meson momentum, the hadronic mixing in lattice units, the energy shift in the charmonium spectrum, and the decay width.}
	\label{tab:results}
\end{table}
\section{Conclusions and outlook} \label{sec:conclusions}

% Summarize results
In this project, we employ the ratio method proposed in \cite{Pennanen:2000yk} to determine the decay width and the energy shift for the process $\HepProcess{\Pgya\to\APD\PD}$, finding the results shown in \cref{tab:results} on two \ac{cls} ensembles with two dynamical flavors. We predict a decay width $\Gamma = \qty{24.2(6.4)}{\mega\eV}$, fully compatible with the experiment. The energy shift, although a theoretical quantity, informs us by how much the spectrum would shift in a fully dynamical simulation. We are able to determine it precisely, and we observe a stronger dependence with the volume than the decay width. We found that the ${}^3P_0$ quark model describes qualitatively the lattice data, providing a physical insight into how to explain the lattice simulation with non-relativistic quantum mechanics. We refer to \cite{neuendorf_23} for further details.

% Novelty of our study
Our study employs \acp{ptbc} on the charm quarks to precisely fix the kinematics, and we solve a \ac{gevp} to isolate the excited charmonium state $\Pgya$. In the literature, the ratio method is commonly used to study transitions between ground states, and we needed to re-derive its basic relations to extract the transition amplitude and account for an excited energy level in the initial state. We found that several relations given in \cite{McNeile:2002az} are no longer applicable in our situation, but enough remains to analyze the lattice data.

% Comparison with the Luscher method
The ratio method relies on the narrow width approximation because, unlike the approach defined in \cite{Luscher:1985dn,Luscher:1986pf,Luscher:1991cf,Luscher:1990ux}, it does not establish a direct connection between the lattice matrix elements and the scattering in infinite volume. This introduces a systematic error difficult to estimate that limits the applicability of this method. Although both approaches rely on good lattice spectroscopy to give a prediction of the decay, the cost is a stark difference between the two methods. Had we done this calculation with \cite{Luscher:1985dn,Luscher:1986pf,Luscher:1991cf,Luscher:1990ux}, we would have considered several volumes and reference frames, having to solve a \ac{gevp} for each of them. Precisely, the high computational costs prevent many works employing \cite{Luscher:1985dn,Luscher:1986pf,Luscher:1991cf,Luscher:1990ux} from taking the continue limit or simulating at physical light-quark masses. And although this is not a fundamental barrier, it continues to be a source of systematic uncertainty.

% Points to improve
The analysis can readily be extended to more ensembles, taking the continuum limit and approaching the physical quark masses. Looking at the difference between ensembles D5 and E5, we observe that the \acp{fve} are smaller than the statistical errors, and we have not considered systematic errors at this stage. Regarding the spectrum, it was shown in \cite{Wilson:2015dqa} that it is important to include in the \ac{gevp} correlators for both the initial and the final states, otherwise there is the possibility to miss some level. In our project, this would require expanding the \ac{gevp} matrix with $\expval*{O^{\APD\PD}O^{\APD\PD,\dagger}}$ and $\expval*{O^{\APD\PD}O^{\psi,\dagger}}$ correlators.

% Outlook
This study can also be expanded modifying the heavy-quark mass. Lowering $\kappa_{\Pcharm}$ towards values closer to $\kappa_{\Pstrange}$, we would approach the transition $\HepProcess{\Pphi\to\PK\APK}$, while increasing $\kappa_{\Pcharm}$ up to $\kappa_{\Pbottom}$, we would draw nearer to $\HepProcess{\PUpsilonFourS\to\PB\APB}$. In practice, we believe that distillation \cite{HadronSpectrum:2009krc} is a prerequisite to correctly isolate the spectrum at a wide variety of heavy-quark masses and reduce the statistical uncertainties. The method presented here could be applied to simulations with more dynamical flavors, although care should be taken that final-state interactions are small to prevent large \acp{fve} due to \acp{ptbc}. In relation to the ${}^3P_0$ quark model, we could perform a comparison with \cite{Segovia:2012cd}, and see if the lattice results in the continuum limit can be described with the functional dependence of $\beta$ given there, which is fitted to the experiment.

\acknowledgments

The work of TSJ was supported by Agence Nationale de la Recherche under the contract ANR-17-CE31-0019, and by the STFC Consolidated Grant ST/X000494/1 Particle Theory at the Higgs Centre. This work is supported in part (J.N. and J.H.) by the program “Netzwerke 2021”, an initiative of the Ministry of Culture and Science of the State of Northrhine Westphalia, in the NRW-FAIR network,
funding code NW21-024-A.
Additionally, this work is supported by the German Research Foundation (DFG) through
the Research Training Group “GRK 2149: Strong and Weak Interactions – from Hadrons to Dark
Matter” (J.N. and J.H.).
This project was granted access to the HPC resources of TGCC (2024-A0160502271) by GENCI.
Parts of the calculations for this publication were performed on the HPC cluster PALMA II of the University of Münster, subsidised by the DFG (INST 211/667-1).
J.H. would like to thank the long-term workshop on HHIQCD2024 at the
Yukawa Institute for Theoretical Physics (YITP-T-24-02) for giving him
a chance for useful discussions as this work was finished.
The authors thank Alain Le Yaouanc and Antonin Portelli for fruitful discussions.

% Acronyms.

\acrodef{apbc}[anti-PBC]{anti-periodic boundary condition}
\acrodef{ape}[APE]{Array Processor Experiment}
\acrodef{cg}[CG]{Clebsch-Gordan}
\acrodef{cls}[CLS]{Coordinated Lattice Simulations}
\acrodef{cm}[CM]{centre-of-mass}
\acrodef{chpt}[ChPT]{chiral perturbation theory}
\acrodef{ddhmc}[DD-HMC]{domain decomposition hybrid Monte Carlo}
\acrodef{fve}[FVE]{finite-volume effect}
\acrodef{gevp}[GEVP]{generalized eigenvalue problem}
\acrodef{lqcd}[LQCD]{lattice QCD}
\acrodef{pbc}[PBC]{periodic boundary condition}
\acrodef{ptbc}[PTBC]{partially twisted boundary condition}
\acrodef{qcd}[QCD]{quantum chromo dynamics}
\acrodef{tbc}[TBC]{twisted boundary condition}

\appendix
\section{\texorpdfstring{${}^3P_0$}{3P0} quark-pair creation model} \label{sec:quark-model}

\begin{figure}
	\centering
	\includegraphics[scale=1]{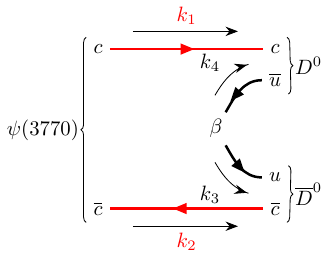}
	\caption[3P0 quark model]{The decay $\HepProcess{\Pgya\to\APD\PD}$ as understood by the ${}^3 P_0$ quark model. The light quarks are created from the vacuum with strength $\beta$. The heavy quarks do not interact, and therefore one should rearrange the quark quantum numbers to obtain the correct final state.}
	\label{fig:3p0-quark-model}
\end{figure}

We would like to describe the dependence on the $\PD$-meson momentum of the matrix element $\braket*{\Pgya}{\APD\PD}$ with an effective theory. In this way, we create a simplified picture of the decay process that we can understand analytically. For this reason, we employ the so-called ${}^3P_0$ quark-pair creation model, which was introduced in \cite{MICU1969521,Carlitz:1970xb} back in 1969 and 1970 to explain $1 \to 2$ particle transitions.
It assumes all hadrons are made out of constituent quarks, and that $\Pcharm$ and $\APcharm$ are mere spectators, such that the final state is obtained creating a quark-antiquark pair from the vacuum with quantum numbers $0^{++}$, see \cref{fig:3p0-quark-model}.
Since the lowest quantum numbers to achieve this are $L=1$ and $S=1$, this results in the configuration ${}^3P_0$ that gives name to the model.
In its most basic form, there is only one free parameter, which we call $\beta$, that gives the strength of the pair creation.
This quark model has been applied to both meson and baryon decays \cite{LeYaouanc:1973ldf,Blundell:1995ev}, including $\psi(4040)$ and $\psi(4415)$ \cite{LeYaouanc:1977fsz,LeYaouanc:1977gm}.
More recently, it has been shown that the agreement with heavy meson decays, including $\Pgya$, can be improved making the coupling $\beta$ a function of the meson reduced mass \cite{Segovia:2012cd}. We will not discuss this work, however, as our analysis employs different conventions and our results are not in the continuum. Besides, there are relativistic implementations \cite{Godfrey:1985xj} that we do not require in this project.
\begin{table}
    \centering
    \begin{tabular}{c|c|c|cc|cc|ccc|ccc}
	   \toprule
	   & $E_0$ & $E_1$ & \multicolumn{2}{c|}{$E_2$} & \multicolumn{2}{c|}{$E_3$} & \multicolumn{3}{c|}{$E_4$} & \multicolumn{3}{c}{$E_5$} \\
	   \midrule
	   n & 0   & 1   & 2   & 2 & 3   & 3 & 4   & 4 & 4 & 5   & 5 & 5 \\
	   l & 0   & 1   & 2   & 0 & 3   & 1 & 4   & 2 & 0 & 5   & 3 & 1 \\
	   \bottomrule
    \end{tabular}
\caption{First few quantum numbers and energies for the harmonic oscillator wave function.}
\label{tab:harmonic-oscillator-quantum-numbers}
\end{table}
We follow the non-relativistic approach by \cite{LeYaouanc:1972vsx}, where the meson wave functions are described by a harmonic oscillator in spherical coordinates. Using natural units, we consider the time-independent solution
\begin{equation}
	\Psi_{nlm}(\bar{p}) = \frac{A_{nl}(p)}{p} Y_l^m(\theta,\phi),
\end{equation}
with the spherical harmonics for the angular dependence and the radial piece
\begin{equation}
	A_{nl}(p) = N(n,l) p^{l+1} e^{-\frac{p^2}{2\gamma}}
	L_{\frac{1}{2}(n-l)}^{(l+1/2)}\left(\frac{p^2}{\gamma}\right).
\end{equation}
We define $\gamma \equiv m\omega$, $L$ is the generalized Laguerre polynomial and the normalization factor is
\begin{equation}
	N(n,l)
	=
	(-\iu)^n\left(\frac{2^{n+l+2}}{\sqrt{\pi}\gamma^{l+3/2}} \frac{\left(\frac{n-l}{2}\right)! \left(\frac{n+l}{2}\right)!}{(n+l+1)!} \right)^{1/2}.
\end{equation}
The quantum number $n$ gives the energy of the oscillator, $E_n=\omega(n+3/2)$, and $l$ and $m$ are the eigenvalues of the total angular momentum and the third component, respectively. Note the quantum numbers must fulfill the following rules: $l$ and $n$ are non-negative integers, with $n \geq l$; $l$ and $n$ must be both even or both odd; and $-l \leq m \leq l$. These rules restrict the combination of quantum numbers to the ones shown in \cref{tab:harmonic-oscillator-quantum-numbers}.

\begin{figure}
	\centering
	\includegraphics[scale=1]{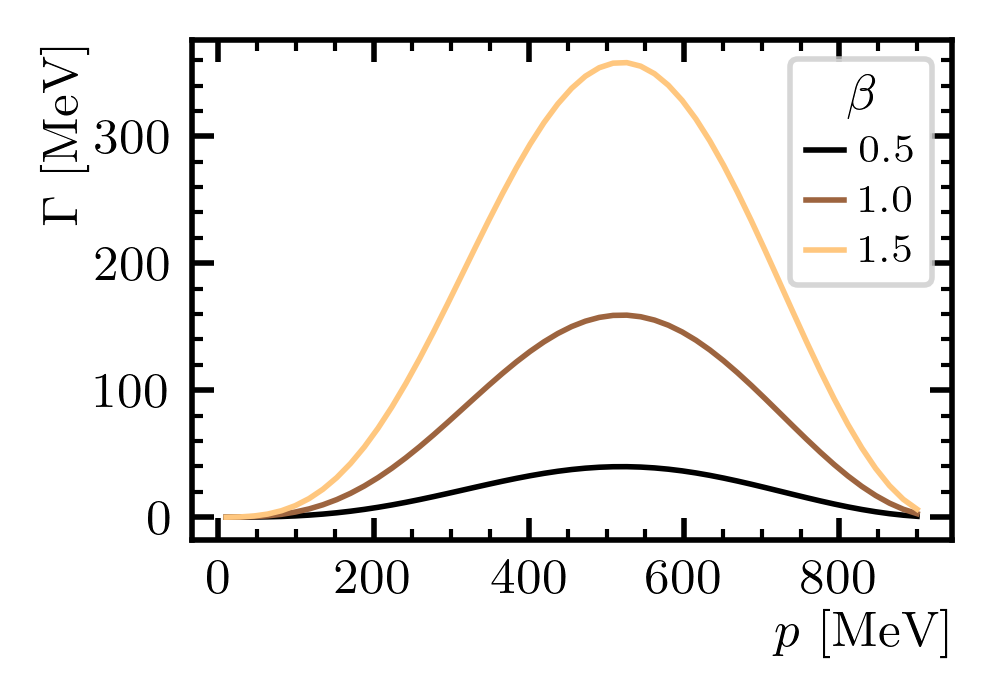}
	\caption[Examples for the decay width from the 3P0 quark model]{The decay width \cref{eq:3p0-decay-width}, given by the physical masses and \cref{eq:3p0-mixing} for the hadronic mixing, as a function of the $\PD$-meson momentum. We plot the function for several values of the quark-pair creation constant $\beta$, which is dimensionless.}
	\label{fig:test_quark_model}
\end{figure}

We assign one wave function to $\Pgya$ and another to the $\PD$-meson. To fully specify them, we need the values of $\omega$ and $m$, which are the oscillator period and the reduced mass, respectively. We also need to identify the physical states, which have particular $J^{\text{PC}}$ quantum numbers, with a given harmonic oscillator. Regarding charmonium, $n=l=0$ corresponds to $\PJpsi$; $n=2$, $l=0$ is $\PpsiTwoS$; and $\Pgya$ is given by $n=l=2$. Note that within the confines of this nonrelativistic description, the spin does not enter explicitly in the wavefunction, and $\PpsiTwoS$ and $\Pgya$ are mass-degenerate. Meanwhile, $n=l=1$ corresponds to $\chi_{c1}(1P)$, which is instead an axial vector. For the $\PD$-meson, we select $n=l=0$.
In addition, we need to provide values for the period of the harmonic oscillator $\omega$ and the mass $m$. The latter corresponds to the reduced mass of the quark-antiquark system of each meson. To analyze our data, we may give values to these quantities based on the lattice results. In this appendix, we use the experimental values as an example: $\omega = \qty{342}{\mega\eV}$ and $\gamma = \qty{92362}{\mega\eV^2}$ for $\PDzero$ and $\APDzero$; $\omega = \qty{294.6}{\mega\eV}$ and $\gamma = \qty{220948.875}{\mega\eV^2}$ for $\Pgya$. The frequency is extracted by looking at the energy difference between consecutive states.
Omitting energy and momentum conservation, the hadronic transition amplitude is \cite{LeYaouanc:1972vsx},
\begin{equation}
    \label{eq:3p0-mixing}
	x_{31}(p) = \frac{1}{\sqrt{30}} I_{0,0} - \frac{1}{2\sqrt{10}}(I_{1,-1} + I_{-1, 1}),
\end{equation}
The prefactors are the appropriate Clebsch-Gordan coefficients and spin matrix elements. The spatial integrals $I_{0,0}$, $I_{1,-1}$ and $I_{-1,1}$ are given by
\begin{equation}
    I_{m,m\prime} = \frac{\beta}{8}\int\dd[3]{k}\mathcal{Y}_1^m(\bar{k}-\bar{k}_{\PD})
    \Psi_{22m\prime}^{\psi}\left(\frac{\bar{k}+\bar{k}_{\PD}}{2}\right)
    \Psi_{000}^{\APD*}\left(\frac{\bar{k}-\bar{k}_{\PD}\alpha}{2}\right)
    \Psi_{000}^{\PD^*}\left(\frac{\bar{k}-\bar{k}_{\PD}\alpha}{2}\right),
\end{equation}
where $\mathcal{Y}$ is a solid harmonic, $k_{\PD}$ is the momentum of a $\PD$-meson in the final state, and $\Psi_{22m\prime}^{\psi}$, $\Psi_{000}^{\APD*}$ and $\Psi_{000}^{\PD^*}$ are the harmonic oscillators associated with $\Pgya$ and $\PD$ respectively. The parameter $\alpha$ is defined in terms of the constituent quark masses, $\alpha \equiv (m_{\Pcharm}-m_{\ell})/(m_{\Pcharm}+m_{\ell})$, where $m_{\Pcharm} \sim \qty{1.5}{\giga\eV}$ and $m_{\ell} \sim \qty{0.33}{\giga\eV}$. Note that $I_{1,-1} = I_{-1,1}$. Since we employ a different normalization for the hadronic transition, we compare the decay width directly with the one obtained from the lattice calculation.
Using \cref{eq:3p0-mixing}, the decay width is \cite{tayduganov:tel-00648217}
\begin{equation}
	\label{eq:3p0-decay-width}
	\Gamma = 8\pi^2 \frac{E^2 p}{m_{\Pgya}} \abs{x_{31}}^2,
    \quad \text{with} \quad
    p^2 = \frac{E^2}{4}-m_{\PD}^2.
\end{equation}
Setting $\bar{k}_{\PD}=(0,0,k_{\PD})$, we plot the decay width as given by the ${}^3P_0$ quark model in \cref{fig:test_quark_model} as a function of the $\PD$-meson momentum and different values of the quark-antiquark coupling, $\beta$. The on-shell condition in experiment, depending on the decay channel, occurs at $p \sim \qty{250}{\mega\eV}$.

% Bibliography

%% [A] Recommended: using JHEP.bst file
\bibliographystyle{JHEP}
\bibliography{biblio.bib}

\end{document}